\documentclass[letter, longauth]{aa} 

\usepackage{float}
\usepackage{graphicx}
\usepackage{multirow,xspace,upgreek}
\usepackage{txfonts}
\newcommand{\mic}{\ensuremath{\upmu}m\xspace}

\newcommand{\Teff}{\ensuremath{\mathrm{T_{eff}}}\xspace}
\newcommand{\Rjup}{\ensuremath{\mathrm{R_{Jup}}}\xspace}
\newcommand{\Mjup}{\ensuremath{\mathrm{M_{Jup}}}\xspace}

\usepackage{natbib}
\usepackage[colorlinks=true,linkcolor=blue,allcolors=blue]{hyperref}

\begin{document} 

  \title{X-SHYNE: X-shooter spectra of young exoplanet analogs\thanks{Based on observations collected at the European Organization for Astronomical Research in the Southern Hemisphere under ESO programme 0101.C-0290.}}

    \subtitle{I. A medium-resolution 0.65--2.5~$\upmu$m one-shot spectrum of VHS\,1256-1257 b}

   \author{S.~Petrus\inst{1, 2}, G.~Chauvin\inst{3}, M.~Bonnefoy\inst{4}, P.~Tremblin\inst{5}, B.~Charnay\inst{6}, P.~Delorme\inst{4}, G.-D.~Marleau\inst{7, 8, 9, 10}, A.~Bayo\inst{11}, E.~Manjavacas\inst{12}, A.-M.~Lagrange\inst{13}, P.~Molli\`ere\inst{10},~P.~Palma-Bifani\inst{3}, B.~Biller\inst{14}, J.-S.~Jenkins\inst{15, 16}
          }


   \institute{
   Instituto de F\'{i}sica y Astronom\'{i}a, Facultad de Ciencias, Universidad de Valpara\'{i}so, Av. Gran Breta\~{n}a 1111, Valpara\'{i}so, Chile\\
                \email{simon.petrus@npf.cl}
   \and
   {N\'{u}cleo Milenio Formac\'{i}on Planetaria - NPF, Universidad de Valpara\'{i}so, Av. Gran Breta\~{n}a 1111, Valpara\'{i}so, Chile}
   \and
   Laboratoire Lagrange, Université Cote d’Azur, CNRS, Observatoire de la Cote d’Azur, 06304 Nice, France
   \and    
   Univ. Grenoble Alpes, CNRS, IPAG, F-38000 Grenoble, France
   \and
   Maison de la Simulation, CEA, CNRS, Univ. Paris-Sud, UVSQ, Université Paris-Saclay, 91191 Gif-sur-Yvette, France
    \and
    LESIA, Observatoire de Paris, Universit\'{e} PSL, CNRS, Sorbonne Universit\'{e}, Univ. Paris Diderot, Sorbonne Paris Cit, 5 place Jules Janssen, 92195 Meudon, France.
    \and
    Fakult\"{a}t f\"{u}r Physik, Universit\"{a}t Duisburg-Essen, Lotharstraße 1, 47057 Duisburg, Germany
    \and
    Institut f\"ur Astronomie und Astrophysik, Universit\"at T\"ubingen, Auf der Morgenstelle 10, D-72076 T\"ubingen, Germany
    \and
    Physikalisches Institut, Universit\"{a}t Bern, Gesellschaftsstr.~6, CH-3012 Bern, Switzerland
     \and
    Max-Planck-Institut f\"{u}r Astronomie, K\"{o}nigstuhl 17, 69117 Heidelberg, Germany
    \and
    European Southern Observatory, Karl Schwarzschild-Stra\ss e 2, D-85748 Garching bei M\"{u}nchen, Germany
    \and
    AURA for the European Space Agency (ESA), ESA Office, Space Telescope Science Institute, 3700 San Martin Drive, Baltimore, MD, 21218 USA
    \and
    LESIA, Observatoire de Paris, Université PSL, CNRS, Sorbonne Université, Université de Paris, 5 place Jules Janssen, 92195 Meudon, France
    \and
    Institute for Astronomy The University of Edinburgh Royal Observatory Blackford Hill Edinburgh EH9 3HJ U.K.
    \and
    N\'ucleo de Astronom\'ia, Facultad de Ingenier\'ia y Ciencias, Universidad Diego Portales, Av. Ej\'ercito 441, Santiago, Chile     \and
     Centro de Astrof\'isica y Tecnolog\'ias Afines (CATA), Casilla 36-D, Santiago, Chile
   }


   \date{Received ---; accepted ---}

 
  \abstract
{We present simultaneous 0.65-2.5~\mic~medium resolution ($3300 \leq R_{\lambda} \leq 8100$) VLT/X-Shooter spectra of the young low-mass ($\mathrm{19\pm5 M_{Jup}}$) L-T transition object VHS 1256-1257 b, a known spectroscopic analogue of HR8799d. The companion is a prime target for the \textit{JWST} Early Release Science (ERS) and one of the highest-amplitude variable brown-dwarf known to date. We compare the spectrum to the custom grids of cloudless ATMO models exploring different atmospheric composition with the Bayesian inference tool \texttt{ForMoSA}. We also re-analyze low-resolution HST/WFC3 1.10-1.67~\mic~ spectra at minimum and maximum variability to contextualize the X-Shooter data interpretation. The models reproduce the slope and most molecular absorption from 1.10 to 2.48~\mic self-consistently but fail to provide a radius consistent with evolutionary model predictions. They do not reproduce consistently the optical spectrum and the depth of the K I doublets in the J-band. We derive \Teff\,=\,1380$\pm$54 K, log(g)\,=\,3.97$\pm$0.48 dex, [M/H]\,=\,0.21$\pm$0.29, and C/O > 0.63. Our inversion of  the HST/WFC3 spectra suggests a relative change of $27^{+6}_{-5}K$ of the disk-integrated \Teff correlated with the near-infrared brightness. Our data  anchor the characterization of that object in the near-infrared and could be used jointly to the ERS mid-infrared data to provide the most detailed characterization of an ultracool dwarf to date.} 


    \titlerunning{X-SHYNE I: A medium-resolution 0.65--2.5~\mic spectrum of VHS\,1256-1257 b}
    \authorrunning{S.\ Petrus, G.\ Chauvin, M.\ Bonnefoy et al.}
 
   \maketitle
%

\section{Introduction}

\begin{figure*}[t]
\centering
\includegraphics[width=\linewidth]{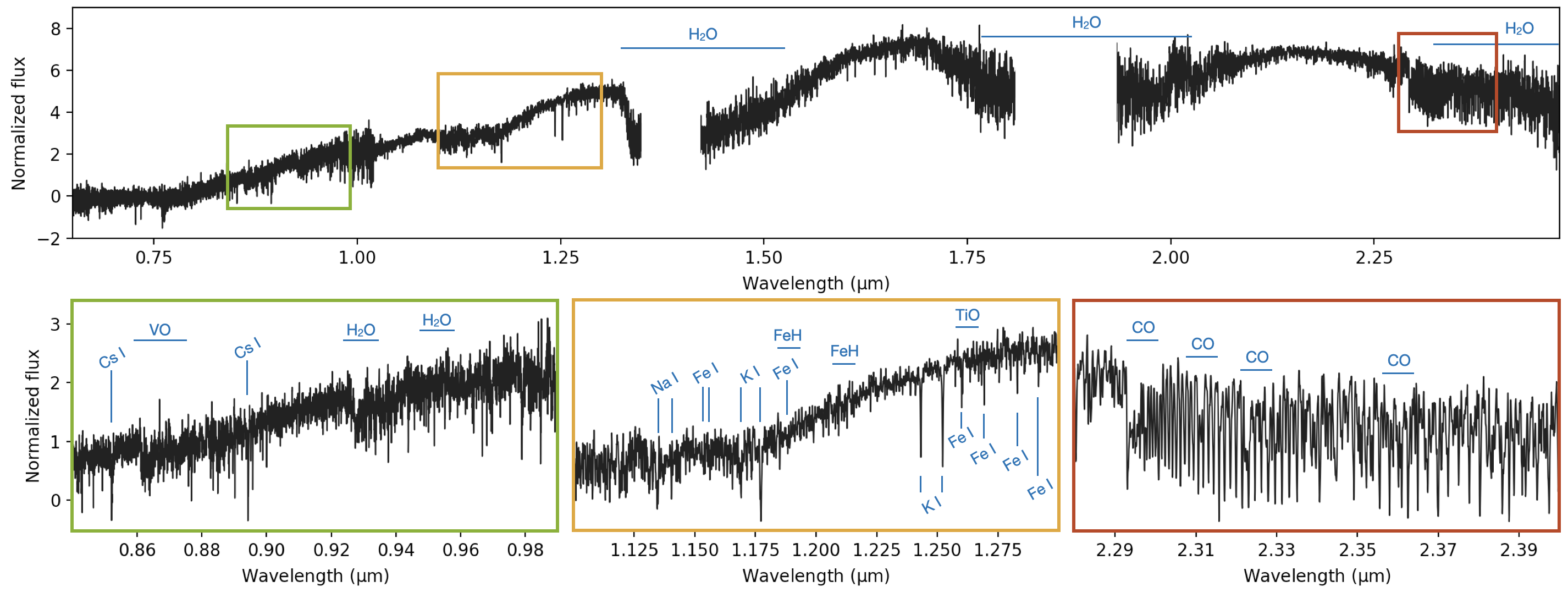}
    \caption{X-shooter optical and near-infrared spectra of VHS\,1256-1257\,b (\textit{black}). Zooms are provided in order to show the atomic and molecular absorptions that are detected in our data.}
    \label{fig:data}
\end{figure*}

 Direct imaging and interferometry can yield high quality spectra of young (5-200 Myr) planets made of tens to thousands of data-points in a few hours of telescope time \citep{2018haex.bookE.101B}. About a dozen of exoplanets have been directly characterized. The majority occupy a slightly distinct space compared to the well defined sequence of mature field M, L, and T brown dwarfs indicative of different physical and atmospheric properties.

Since 2001, a few tens of free-floating objects sharing the same age and mass range as imaged exoplanets have also been identified in young star forming regions \citep{2001MNRAS.326..695L} and more recently in young nearby associations where most imaged exoplanets are identified \citep[e.g.][and ref. therein]{2016ApJS..225...10F}. They are seconded by a population of ultrawide orbit ($\geq100$\,au) companions to low-mass stars sharing the same properties \citep[e.g.,][]{2008ApJ...673L.185B, 2014ApJ...787....5N, 2015ApJ...806..254A}.  Contrary to imaged planets blurred into the stellar halo, these objects can be characterized at high signal-to-noise (S/N) with seeing-limited spectrographs over a broad spectral range. They have been found to be precious empirical templates to imaged exoplanets having similar placements in color-magnitude diagrams and spectral features at low-resolving powers  \citep[e.g.,][]{2010A&A...512A..52B, 2017AJ....153..182C, 2017A&A...605L...9C}.

The so-called L--T transition is evidenced by a sharp bluering of near-infrared colors thought to appear at nearly-constant effective temperatures in mature field dwarfs \citep{2015ApJ...810..158F}. The transition appears to be even more extreme for young objects, which are redder and underluminous by up to a few magnitudes compared to mature field dwarf counterparts. Sophisticated models of clouds of different compositions (silicates, sulfites...) have been considered for more than a decade to establish the basis of our understanding of atmospheric physics and reproduce the collected spectra. Although they manage to reproduce the spectra of massive late-M to early-L type young brown dwarfs and exoplanets well, they fail to fit the luminosity and the spectra of cooler L--T transition planets \citep{2016A&A...587A..58B}. The problem likely arises from uncertainties in the cloud model itself in terms of composition, particle size and density, and also possible cloud deck inhomogeneity.  The modification of the cloud vertical and cloud particle  distribution is corroborated by recent variability studies that show higher-amplitude variability in young L dwarfs than old ones \citep{2015ApJ...799..154M, 2018AJ....155...11M}. Additional processes such as desequilibrium chemistry  \citep{2014ApJ...792...17S} and related thermo-chemical instability \citep{2015ApJ...804L..17T} are now also included in some models. The  resulting warming up of the deep atmosphere along the L--T transition is proposed to explain the properties of these objects, including their rotational spectral modulations \citep{2020A&A...643A..23T}. However, today’s low-resolution spectra of imaged exoplanets covering a small portion of the planet spectral energy distribution do not allow to explore and remove the atmospheric models degeneracies. 
 

In 2018, we initiated the \textit{X-SHooter spectra of YouNg Exoplanet analogs} program (X-SHYNE) to acquire simultaneous medium-resolution 0.3-2.5~\mic~ spectra of forty young M, L, and T-type exoplanet analogs. Medium spectral resolution enables a better deblending of the atomic and the molecular lines  and offer the rich perspective to untangle temperature, age, atmospheric composition, cloud properties, and disequilibrium chemistry in the spectra and improve the current generation of planetary atmosphere models. 

In this letter, we focus on the young L--T transition object VHS J125601.92-125723.9\,b (hereafter VHS\,1256-1257\,b; see Table\,\ref{tab:para_sys}). This companion was detected at $8.06\pm0.03$\,as (projected physical separation of 179$\pm$9 au) from the tight 0.1\,as equal-magnitude late-M binary VHS\,J125601.92-125723.9\,AB \citep{gauza2015, 2016ApJ...818L..12S}. VHS 1256-1257 b is an excellent spectral analogue of the emblematic exoplanet HR8799d \citep{2016A&A...587A..58B} and one of the first target observed with the \textit{James Webb Space Telescope} (\textit{JWST}) which  provides 3 to 28~\mic~medium-resolution spectra complementary to the X-SHYNE data \citep{2022arXiv220512972H}.  In this study, we present in Section \ref{section:2} the X-SHYNE observations and the data reduction of VHS\,1256-1257\,b. In Section \ref{section:3}, we describe our empirical study and the forward modeling analysis of the companion spectrum. In Section \ref{section:4}, we discuss  atmospheric properties and prospects for future synergy and interpretation  with the \textit{JWST} data. 

\begin{table}[t!]
   \centering
    \small
         \caption{Physical properties of VHS\,J125601.92-125723.9\,AB and b.} 
    \begin{tabular}{lccc}
       \hline\hline\noalign{\smallskip}
        Properties & Binary & Companion  & Ref. \\
         \noalign{\smallskip}\hline\noalign{\smallskip}
         Spectral type & M$7.5\pm0.5$  & L7$\pm$1.5    & 1      \\
         \noalign{\smallskip}Luminosity ($\log$\,L$_\odot$)    & $-2.95\pm0.07$  & $-4.54\pm0.07$  & 2 \\
         \noalign{\smallskip}Teff (K)        & $2902_{-53}^{88}$ &  $1240\pm50$ & 2, 3 \\
         \noalign{\smallskip}log(g)  (dex)        & $5.22_{-0.02}^{+0.01}$   &  $4.55_{-0.11}^{+0.15}$ & 2, 3  \\
         \noalign{\smallskip}Mass  (\Mjup)         & $114_{-5}^{+7}$ & $19\pm5$   & 2, 3 \\
         \noalign{\smallskip}Radius  (\Rjup)         & $1.33_{-0.05}^{+0.07}$ & $1.17\pm0.04$ & 2, 3    \\
         \noalign{\smallskip}         Distance (pc)      & \multicolumn{2}{c}{$22.2_{-1.2}^{+1.1}$} & 2\\
         \noalign{\smallskip}Age (Myr)            & \multicolumn{2}{c}{150--300} & 1, 2 \\
        \noalign{\smallskip}Projected sep. (as) &   0.109$\pm$0.002  & 8.06$\pm$0.03 & 3   \\
       \noalign{\smallskip}Physical sep. (au) &   2.4$\pm$0.1 & 179$\pm$9 & 2  \\
         \noalign{\smallskip}\hline \noalign{\smallskip}\noalign{\smallskip}        
             \end{tabular} \quad
    \label{tab:para_sys}
\tablefoot{The companion \Teff, mass, radius, and surface gravity are updated from evolutionary model predictions \citep{2003A&A...402..701B}. References: 1 - \cite{gauza2015}, 2 - \cite{2020RNAAS...4...54D}, 3 - this work.}
\end{table}

\section{Observation and data reduction}
\label{section:2}

VHS\,1256-1257\,b was observed on May 28th, 2018 with the X-shooter seeing-limited medium-resolution spectrograph mounted at the VLT/UT2 Cassegrain focus \citep{2011A&A...536A.105V}.  The wide wavelength coverage of the instrument (0.300-2.480~\mic) obtained at one shot prevents from uncertainties arising from the collage of spectra of this variable object obtained on separate bands at different nights. We chose the 1.6\arcsec, 1.2\arcsec and 0.6\arcsec-wide slits for the UVB, VIS, and NIR arms corresponding to resolving powers $R_{\lambda}$ = $\lambda/\Delta\lambda$ = 3300, 6700, and 8100, respectively. The slits were oriented perpendicular to the position angles of the companion in order to mitigate the flux contamination of the host stars. The target was observed following an ABBA strategy to evaluate and remove the sky emission during the data processing step. The observing conditions and details of the observing log are reported in Appendix \ref{Appendix:obs_log}.  We used the ESO \texttt{reflex}  framework \citep{2013A&A...559A..96F} to run the X-shooter pipeline version 2.9.3 on the raw data \citep{2010SPIE.7737E..28M}. The pipeline produces two-dimensional, curvature-corrected, and flux-calibrated spectra for each target and epoch of observation (trace). The spectra were extracted from the traces using a custom python script. The flux in each wavelength channel at the position of the source was averaged within 720\,mas aperture in the UVB and VIS arms, and a 1120\,mas aperture in the NIR arm. The script computed the noise at the position of the source into each spectral channel following the procedure described in \cite{2020AA...633A.124P}. The residual nonlinear pixels in the spectra were removed using the kappa-sigma clipping method. The telluric corrections were evaluated and removed using the \texttt{molecfit} package \citep{2015A&A...576A..77S,2015A&A...576A..78K}. The spectra at each epoch were corrected from the barycentric velocity and flux-calibrated  using a spectro-photometric standard. Finally, we merged our two spectra in one spectrum. 

\section{Results}
\label{section:3}

The X-shooter extracted spectrum of VHS\,1256-1257\,b is reported in Fig.\,\ref{fig:data}. We decided to exclude data shorter than 0.65~\mic~ due to the low S/N at these wavelengths. We performed an empirical analysis that confirmed the young age of this target (see Appendix \ref{App:emp_ana}). 


\begin{figure*}[h!]
\centering
\includegraphics[width=1\hsize]{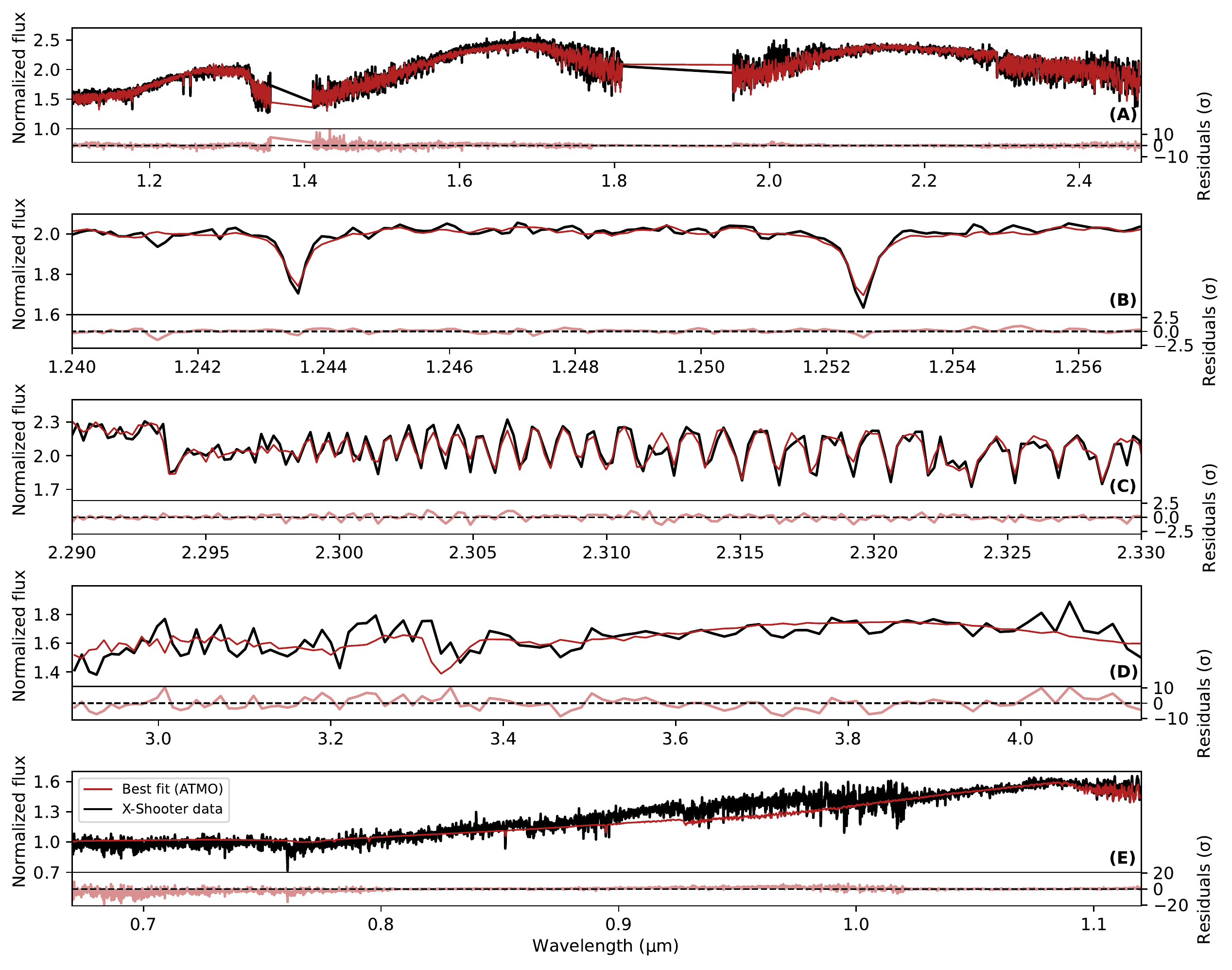}
    \caption{X-shooter spectra of VHS\,1256-1257\,b (\textit{black}) compared to the best fit \texttt{ATMO} model (\textit{red}). Residuals from the best fit are shown at the bottom of each panel. Fit between \textbf{(A)}:1.10 and 2.48~\mic; \textbf{(B)}: 1.225 and 1.275~\mic; \textbf{(C)}: 2.28 and 2.41~\mic; \textbf{(D)}: 2.90 and 4.14~\mic. \textbf{(E)}: 0.65-2.48~\mic~ fit evidencing the model departure at optical wavelengths.}
    \label{fig:FITS}
\end{figure*}

To explore the spectral diversity of the photometric and spectroscopic observations, we used \texttt{ForMoSA}, a tool based on a forward-modelling approach that compares observations with grids of pre-computed synthetic atmospheric models using Bayesian inference methods. This code is presented in \cite{2020AA...633A.124P} and \cite{2021AA...648A..59P} but has been updated for this study.The grids of synthetic spectra used as input are now provided in the standardized NetCDF4 format through the use of the  \texttt{xarray}\footnote{xarray.pydata.org/en/stable/} module, created and maintained by the community of meteorologists and geophysicists, which allows the labeling of N-dimensional grids, as well as parallel computations on these grids (interpolations, extrapolations, and arithmetic). 

We used the grids of synthetic spectra produced by the last generation of the \texttt{ATMO} \citep{2015ApJ...804L..17T} models. Its specificities and the parameter space that it explores are described in the Appendix \ref{Appendix:grids}. 
To limit the impact of possible bias on our estimates of the atmospheric parameters, we applied the strategy described by \cite{2020AA...633A.124P} which consists in defining different optimal wavelength ranges for different adjustments according to the specific parameter studied. Thus, to estimate the \Teff~and the radius, we have performed a fit between 1.10 and 2.48~\mic, masking out the optical part of the spectrum that is not well reproduced by the models (panel (E) of Figure \ref{fig:FITS}). The radius is estimated using the dilution factor $C_{k} = (R/d)^{2}$, with $R$ the target's radius and $d$ the distance. To estimate the surface gravity, the metallicity, the radial velocity, and the rotational velocity, we have chosen to take advantage of the absorption doublet K\,I with the higher S/N which is sensitive to these parameters. We therefore fit our data between 1.225 and 1.275~\mic, without the continuum, to avoid the contribution of the pseudo-continuum and molecular absorption to the fit. We also fit our data between 2.28 and 2.41~\mic~to exploit the CO overtones, particularly sensitive to C/O. For these two last configurations, the dilution factor $C_{k}$ is calculated analytically using the relation of \cite{2008ApJ...678.1372C}. These different fits are shown on Figure\,\ref{fig:FITS}. We also performed independent fits on the $J-$, $H-$, $K-$band, in addition to the $L_{P}$ spectrum of VHS 1256-12257 b from \citealt{2018ApJ...869...18M}. The adopted parameters from each run is given in the Table \ref{tab:fits_param}.



\section{Discussion}
\label{section:4}


In the context of the discovery of VHS\,1256-1257\,b, \cite{gauza2015} led a first empirical characterization to derive the spectral type, but also the bulk properties of the companion.
Confronting the luminosity of VHS\,1256-1257\,b, derived from the near-infrared photometry, bolometric correction and first parallax estimate ($\pi = 78.8 \pm 6.4$\,mas, $12.7\pm1.0$\,pc), to the predictions of the \texttt{BT-SETTL} evolutionary models, they derived physical values for the luminosity, effective temperature, surface gravity and mass of: $L = -5.05\pm0.22$ dex, \Teff\,=\,$880^{+140}_{-110}$\,K, log(g)\,=\,$4.25^{+0.35}_{-0.10}$, and M\,=\,$11.2^{+1.8}_{-9.7}$\,M$_{\rm Jup}$, respectively. Their solutions led to a rather cool atmosphere and unusually under-luminous object given its spectral type. The parallax measurement was recently revised by \cite{2020RNAAS...4...54D} based on CFHT and Pan-STARRS observations placing this system at larger distances: $51.6\pm3.0$\,mas ($22.2^{+1.1}_{-1.2}$\,pc). This new estimate implies higher values for the mass ($19\pm5$\,M$_{\rm Jup}$), surface gravity (log(g) = $4.55^{+0.15}_{-0.11}$), radius ($1.17^{+0.04}_{-0.04}$~\Rjup), and warmer temperature ($1240\pm50$\,K) from the \cite{2008ApJ...689.1327S} hybrid evolutionary models. It brings the absolute magnitudes of all three components in better agreement with known young objects. From a first atmospheric model analysis using near-infared spectra (NTT/SofI, 0.95-2.52\,\mic, $R_{\lambda}$\,=\,600) and L spectra (Keck/NIRSPEC at $L_{P}$-band, $R_{\lambda}$\,=\,1300), \cite{2018ApJ...869...18M} had already derived a warmer temperature of 1240\,K with photospheric clouds for VHS\,1256-1257\,b although surface gravity and radius estimates were still affected by the early, wrong distance determination. From their $L_{P}$-band spectra, \cite{2018ApJ...869...18M} also reported the detection of low abundances of CH$_{4}$, suggesting non-equilibrium chemistry processes at play between CO and CH$_{4}$. The upper atmospheres of their best-fit models depart from equilibrium abundances of CH$_{4}$ by factors of 10–100.

Our atmospheric forward-modeling analysis of the X-shooter spectra confirms the temperature of $1326\pm2$\,K with \texttt{ATMO} models for a full fitting applied between 1.10 and 2.48\,\mic. Nevertheless, this estimate varies significantly with the choice of the wavelength ranges used for the fit from $1145\pm13$\,K for the $L_{P}$-band to $1417\pm17$\,K for the $K$-band.
For the surface gravity determination, we focused on the fitting of the gravity-sensitive K\,I lines between 1.225-1.275\,\mic. The resulting solution indicates surface gravity of log(g)\,=\,$4.25\pm0.20$ dex. But this estimate tends to be lower (< 4.0 dex) when the fit is performed on a larger wavelength range ($J$-, $H$-, $K$-bands).
Although, effective temperature and surface gravity are relatively consistent with the evolutionary model, the radius estimate (R between 0.81 and 0.91~\Rjup) given by the \texttt{ATMO} models is not, and definitively needs to be further investigated for this family of models. At the moment, pressure levels that are impacted by the reduced temperature gradient is assumed to be a function of surface gravity (see Appendix \ref{Appendix:grids} for details), an adjustment of the size of the layer impacted by the reduced temperature gradient might be needed to get a radius in better agreement with evolutionary models. We notice also that the radius estimated from the $L_{P}$-band is consistent with the evolutionary model but the surface gravity is inconsistent with that obtained on the other band and too low with respect to evolutionary model predictions. Finally, accessing spectral resolution up to 8100 at $K$-band, for which water and carbon monoxide are detected, enables us to constrain the C/O ratio between solar and super-solar values with C/O $\ge 0.63$. The metallicity appears to be solar to super-solar for all the fits (except at $K$-band and $H$-band). In the context of the cloudless \texttt{ATMO} models, the adiabatic index value of $\gamma = 1.02\pm0.01$ is consistent with a decrease of the adiabatic index in the lower part of the atmospheres, and processes of non-equilibrium chemistry and quench CO/CH$_{4}$ (i.e. prevent the formation of CH$_{4}$), as expected for young late-L dwarfs subject in this scenario to a CO/CH$_{4}$ thermo-chemical instability \cite{2016ApJ...817L..19T}. Finally, we also derived radial and rotational velocities of RV between -4.58 and 8.77 km\,s$^{-1}$, and v\,sin(i) < 37\,km\,s$^{-1}$ (limited by the spectral resolution of X-shooter), compatible with the NIRSPEC data (RV = $2.1^{+1.6}_{-1.7}$\,km\,s$^{-1}$ and v\,sin(i) = $13.5^{+3.6}_{-4.1}$\,km\,s$^{-1}$, \citealt{2018NatAs...2..138B}). 

\begin{figure*}[!t]
\centering
\includegraphics[width=0.95\hsize]{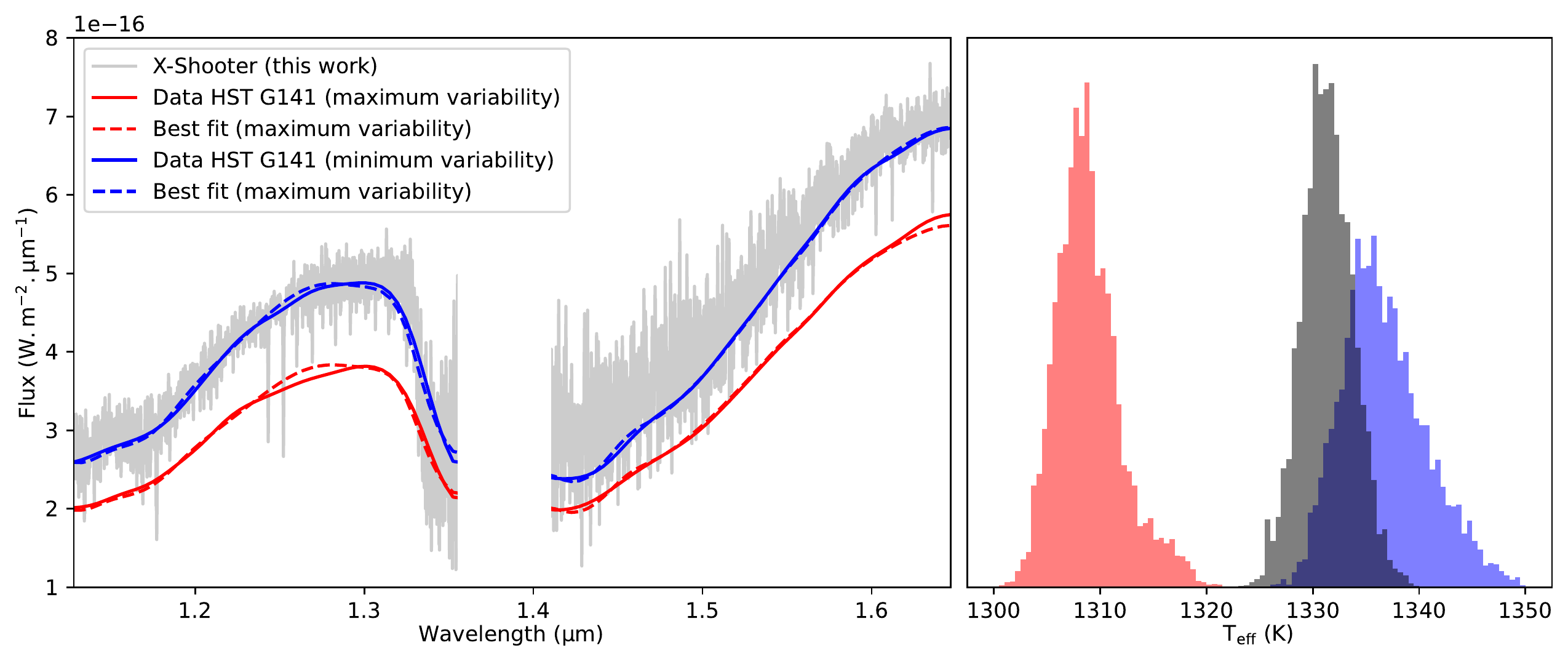}
    \caption{\textbf{Left}: \texttt{ForMoSA} fits (\textit{dash lines}) of HST G141 data (\textit{solid lines}). We used the maximum (\textit{blue}) and minimum (\textit{red}) amplitude data from the sequence of \cite{2020ApJ...893L..30B}. We also compared these HST data with our X-Shooter data. \textbf{Right}: Comparison between \Teff~posteriors from the HST's fits (\textit{red} and \textit{blue}) and the fit of our X-shooter data (\textit{gray}).}
    \label{fig:comp_HST}
\end{figure*}
The [M/H] and the C/O ratio that we measured are very similar to those measured for HR8799 b (C/O = $0.61^{+0.09}_{-0.03}$, \citealt{2015ApJ...804...61B}), HR8799 c (C/O = $0.65^{+0.10}_{-0.05}$, \citealt{2013Sci...339.1398K}), and HR8799 e ([M/H] = $0.48^{+0.25}_{-0.29}$ and C/O = $0.60^{+0.07}_{-0.08}$, \citealt{2020A&A...640A.131M}). The solar to super-solar composition of VHS\,1256-1257\,b could indicate a specific atmospheric metal enrichment of solids and gas during its phases of formation and evolution if formed within a disk by core accretion or even gravitational instability \citep{2022arXiv220413714M}. However, the existence of such a massive planet-forming disk, as well as the presence of giant planets, is not expected given the very-low mass of the central binary \citep{2021A&A...656A..72B}. In addition, the hierarchical architecture of the whole system, a tight equal-mass binary with a planetary-mass companion orbiting at wide orbit, strongly supports a stellar-like formation by gravo-turbulent fragmentation. Further spectroscopic observations to directly measure of the atmospheric composition of the central binary will be important. More extended atmosphere modeling will be crucial to confirm the composition of VHS\,1256-1257\,b, but also its photometric temporal variability. %

VHS\,1256-1257\,b has been identified as one of the most variable substellar object (19.3\% spectroscopic variations between 1.1 and 1.7~\mic) by \cite{2020ApJ...893L..30B}, thanks to a sequence of 11 observations (8.5\,hr) performed with \textit{HST}. The presence of inhomogeneous cloud covers has been proposed and explored for this companion by \cite{2020AJ....160...77Z}, but the existence of temperature fluctuations arising in a convective atmosphere could be an alternative explanation \citep{2020A&A...643A..23T}.
To explore the impact of this spectroscopic variability, we applied \texttt{ForMoSA} using the \texttt{ATMO} models on the faintest and brightest state of this spectroscopic sequence to estimate the \Teff~and compared them with the \Teff~estimated from our X-shooter data. For these fits, we calculated the dilution factor $C_{k}$ analytically and calculated the likelihood using the same wavelength range. The Figure \ref{fig:comp_HST} illustrates the fits of the \textit{HST} data (\textit{Left} panel) and the corresponding \Teff~posteriors (\textit{Right} panel). The difference of temperatures between the faintest and brightest states could be indicative of a temperature fluctuation in the atmosphere of VHS\,1256-1257\,b. 
Modulations of the temperature gradient in the region around one bar could also explain the possible CH$_4$ variability detection observed by \citep{2020ApJ...893L..30B}. Finally, we note also that the \Teff~estimate from our X-shooter data, using the same wavelength range than the \textit{HST} data (gray posterior), is coherent with the \Teff~estimate with the brightest state of the \textit{HST} spectra. 


VHS\,1256-1257\,AB and b is a prime target of the \textit{JWST} Early Release Science (ERS) program (\citealt{2022arXiv220512972H}, PI: S.\ Hinkey) that acquired high-fidelity spectra with NIRSpec and MIRI at medium spectral resolution over a broad wavelength range from 1 to 28~\mic (R$_{\lambda}\sim1500$--3000), respectively. The characterization of these spectra using the forward modeling approach will be treated in a future paper. This will be particularly interesting to accurately constrain the effective temperature of VHS1256\,b, the abundances of important secondary gases (CO, H$_{2}$O, CH$_{4}$), the importance of thermo-
chemical instability, patchy cloud decks, and non-equilibrium chemistry processes at play. As proposed by \cite{2020A&A...643A..23T}, the detection of direct cloud spectral signatures like the silicate absorption feature at 10\,\mic, as detected for ultracool dwarfs observed with \textit{Spitzer} \citep{2022MNRAS.513.5701S}, will help to confirm the presence
of clouds and their contribution to the spectral energy distribution and modulation of VHS\,1256-1257\,b. The ERS program will serve as benchmark for future characterization studies of young brown dwarfs and exoplanetary atmospheres. 




\begin{acknowledgements}
We would like to thank the staff of ESO VLT for their support at the telescope at Paranal and La Silla, and the preparation of the observation at Garching. This publication made use of the SIMBAD and VizieR database operated at the CDS, Strasbourg, France.  This work has made use of data from the European Space Agency (ESA) mission Gaia (https://www.cosmos.esa.int/gaia), processed by the Gaia Data Processing and Analysis Consortium (DPAC, https://www.cosmos.esa.int/web/gaia/dpac/consortium). Funding for the DPAC has been provided by national institutions, in particular the institutions participating in the Gaia Multilateral Agreement.  We acknowledge support in France from the French National Research Agency (ANR) through project grants ANR-14-CE33-0018 and ANR-20-CE31-0012.  
G-DM acknowledges the support of the DFG priority program SPP 1992 ``Exploring the Diversity of Extrasolar Planets'' (MA~9185/1) and from the Swiss National Science Foundation under grant 200021\_204847 ``PlanetsInTime''.
Parts of this work have been carried out within the framework of the NCCR PlanetS supported by the Swiss National Science Foundation.
\end{acknowledgements}

%
%

\bibliographystyle{aa} 
\bibliography{VHS1256b_XSHYNE}

\begin{thebibliography}{48}
\expandafter\ifx\csname natexlab\endcsname\relax\def\natexlab#1{#1}\fi

\bibitem[{{Allers} \& {Liu}(2013)}]{2013ApJ...772...79A}
{Allers}, K.~N. \& {Liu}, M.~C. 2013, \apj, 772, 79

\bibitem[{{Artigau} {et~al.}(2015){Artigau}, {Gagn{\'e}}, {Faherty}, {Malo},
  {Naud}, {Doyon}, {Lafreni{\`e}re}, \& {Beletsky}}]{2015ApJ...806..254A}
{Artigau}, {\'E}., {Gagn{\'e}}, J., {Faherty}, J., {et~al.} 2015, \apj, 806,
  254

\bibitem[{{Baraffe} {et~al.}(2003){Baraffe}, {Chabrier}, {Barman}, {Allard}, \&
  {Hauschildt}}]{2003A&A...402..701B}
{Baraffe}, I., {Chabrier}, G., {Barman}, T.~S., {Allard}, F., \& {Hauschildt},
  P.~H. 2003, \aap, 402, 701

\bibitem[{{Barman} {et~al.}(2015){Barman}, {Konopacky}, {Macintosh}, \&
  {Marois}}]{2015ApJ...804...61B}
{Barman}, T.~S., {Konopacky}, Q.~M., {Macintosh}, B., \& {Marois}, C. 2015,
  \apj, 804, 61

\bibitem[{{B{\'e}jar} {et~al.}(2008){B{\'e}jar}, {Zapatero Osorio},
  {P{\'e}rez-Garrido}, {{\'A}lvarez}, {Mart{\'\i}n}, {Rebolo},
  {Vill{\'o}-P{\'e}rez}, \& {D{\'\i}az-S{\'a}nchez}}]{2008ApJ...673L.185B}
{B{\'e}jar}, V.~J.~S., {Zapatero Osorio}, M.~R., {P{\'e}rez-Garrido}, A.,
  {et~al.} 2008, \apjl, 673, L185

\bibitem[{{Biller} \& {Bonnefoy}(2018)}]{2018haex.bookE.101B}
{Biller}, B.~A. \& {Bonnefoy}, M. 2018, in Handbook of Exoplanets, ed. H.~J.
  {Deeg} \& J.~A. {Belmonte}, 101

\bibitem[{{Bonnefoy} {et~al.}(2014){Bonnefoy}, {Chauvin}, {Lagrange}, {Rojo},
  {Allard}, {Pinte}, {Dumas}, \& {Homeier}}]{2014AA...562A.127B}
{Bonnefoy}, M., {Chauvin}, G., {Lagrange}, A.~M., {et~al.} 2014, \aap, 562,
  A127

\bibitem[{{Bonnefoy} {et~al.}(2010){Bonnefoy}, {Chauvin}, {Rojo}, {Allard},
  {Lagrange}, {Homeier}, {Dumas}, \& {Beuzit}}]{2010A&A...512A..52B}
{Bonnefoy}, M., {Chauvin}, G., {Rojo}, P., {et~al.} 2010, \aap, 512, A52

\bibitem[{{Bonnefoy} {et~al.}(2016){Bonnefoy}, {Zurlo}, {Baudino}, {Lucas},
  {Mesa}, {Maire}, {Vigan}, {Galicher}, {Homeier}, {Marocco}, {Gratton},
  {Chauvin}, {Allard}, {Desidera}, {Kasper}, {Moutou}, {Lagrange}, {Antichi},
  {Baruffolo}, {Baudrand}, {Beuzit}, {Boccaletti}, {Cantalloube}, {Carbillet},
  {Charton}, {Claudi}, {Costille}, {Dohlen}, {Dominik}, {Fantinel},
  {Feautrier}, {Feldt}, {Fusco}, {Gigan}, {Girard}, {Gluck}, {Gry}, {Henning},
  {Janson}, {Langlois}, {Madec}, {Magnard}, {Maurel}, {Mawet}, {Meyer},
  {Milli}, {Moeller-Nilsson}, {Mouillet}, {Pavlov}, {Perret}, {Pujet}, {Quanz},
  {Rochat}, {Rousset}, {Roux}, {Salasnich}, {Salter}, {Sauvage}, {Schmid},
  {Sevin}, {Soenke}, {Stadler}, {Turatto}, {Udry}, {Vakili}, {Wahhaj}, \&
  {Wildi}}]{2016A&A...587A..58B}
{Bonnefoy}, M., {Zurlo}, A., {Baudino}, J.~L., {et~al.} 2016, \aap, 587, A58

\bibitem[{{Bowler} {et~al.}(2020){Bowler}, {Zhou}, {Morley}, {Kataria},
  {Bryan}, {Benneke}, \& {Batygin}}]{2020ApJ...893L..30B}
{Bowler}, B.~P., {Zhou}, Y., {Morley}, C.~V., {et~al.} 2020, \apjl, 893, L30

\bibitem[{{Bryan} {et~al.}(2018){Bryan}, {Benneke}, {Knutson}, {Batygin}, \&
  {Bowler}}]{2018NatAs...2..138B}
{Bryan}, M.~L., {Benneke}, B., {Knutson}, H.~A., {Batygin}, K., \& {Bowler},
  B.~P. 2018, Nature Astronomy, 2, 138

\bibitem[{{Burn} {et~al.}(2021){Burn}, {Schlecker}, {Mordasini}, {Emsenhuber},
  {Alibert}, {Henning}, {Klahr}, \& {Benz}}]{2021A&A...656A..72B}
{Burn}, R., {Schlecker}, M., {Mordasini}, C., {et~al.} 2021, \aap, 656, A72

\bibitem[{{Chauvin} {et~al.}(2017){Chauvin}, {Desidera}, {Lagrange}, {Vigan},
  {Gratton}, {Langlois}, {Bonnefoy}, {Beuzit}, {Feldt}, {Mouillet}, {Meyer},
  {Cheetham}, {Biller}, {Boccaletti}, {D'Orazi}, {Galicher}, {Hagelberg},
  {Maire}, {Mesa}, {Olofsson}, {Samland}, {Schmidt}, {Sissa}, {Bonavita},
  {Charnay}, {Cudel}, {Daemgen}, {Delorme}, {Janin-Potiron}, {Janson},
  {Keppler}, {Le Coroller}, {Ligi}, {Marleau}, {Messina}, {Molli{\`e}re},
  {Mordasini}, {M{\"u}ller}, {Peretti}, {Perrot}, {Rodet}, {Rouan}, {Zurlo},
  {Dominik}, {Henning}, {Menard}, {Schmid}, {Turatto}, {Udry}, {Vakili}, {Abe},
  {Antichi}, {Baruffolo}, {Baudoz}, {Baudrand}, {Blanchard}, {Bazzon}, {Buey},
  {Carbillet}, {Carle}, {Charton}, {Cascone}, {Claudi}, {Costille}, {Deboulbe},
  {De Caprio}, {Dohlen}, {Fantinel}, {Feautrier}, {Fusco}, {Gigan}, {Giro},
  {Gisler}, {Gluck}, {Hubin}, {Hugot}, {Jaquet}, {Kasper}, {Madec}, {Magnard},
  {Martinez}, {Maurel}, {Le Mignant}, {M{\"o}ller-Nilsson}, {Llored}, {Moulin},
  {Orign{\'e}}, {Pavlov}, {Perret}, {Petit}, {Pragt}, {Puget}, {Rabou},
  {Ramos}, {Rigal}, {Rochat}, {Roelfsema}, {Rousset}, {Roux}, {Salasnich},
  {Sauvage}, {Sevin}, {Soenke}, {Stadler}, {Suarez}, {Weber}, {Wildi},
  {Antoniucci}, {Augereau}, {Baudino}, {Brandner}, {Engler}, {Girard}, {Gry},
  {Kral}, {Kopytova}, {Lagadec}, {Milli}, {Moutou}, {Schlieder},
  {Szul{\'a}gyi}, {Thalmann}, \& {Wahhaj}}]{2017A&A...605L...9C}
{Chauvin}, G., {Desidera}, S., {Lagrange}, A.~M., {et~al.} 2017, \aap, 605, L9

\bibitem[{{Chilcote} {et~al.}(2017){Chilcote}, {Pueyo}, {De Rosa}, {Vargas},
  {Macintosh}, {Bailey}, {Barman}, {Bauman}, {Bruzzone}, {Bulger}, {Burrows},
  {Cardwell}, {Chen}, {Cotten}, {Dillon}, {Doyon}, {Draper}, {Duch{\^e}ne},
  {Dunn}, {Erikson}, {Fitzgerald}, {Follette}, {Gavel}, {Goodsell}, {Graham},
  {Greenbaum}, {Hartung}, {Hibon}, {Hung}, {Ingraham}, {Kalas}, {Konopacky},
  {Larkin}, {Maire}, {Marchis}, {Marley}, {Marois}, {Metchev},
  {Millar-Blanchaer}, {Morzinski}, {Nielsen}, {Norton}, {Oppenheimer},
  {Palmer}, {Patience}, {Perrin}, {Poyneer}, {Rajan}, {Rameau},
  {Rantakyr{\"o}}, {Sadakuni}, {Saddlemyer}, {Savransky}, {Schneider}, {Serio},
  {Sivaramakrishnan}, {Song}, {Soummer}, {Thomas}, {Wallace}, {Wang},
  {Ward-Duong}, {Wiktorowicz}, \& {Wolff}}]{2017AJ....153..182C}
{Chilcote}, J., {Pueyo}, L., {De Rosa}, R.~J., {et~al.} 2017, \aj, 153, 182

\bibitem[{{Cushing} {et~al.}(2008){Cushing}, {Marley}, {Saumon}, {Kelly},
  {Vacca}, {Rayner}, {Freedman}, {Lodders}, \& {Roellig}}]{2008ApJ...678.1372C}
{Cushing}, M.~C., {Marley}, M.~S., {Saumon}, D., {et~al.} 2008, \apj, 678, 1372

\bibitem[{{Cushing} {et~al.}(2005){Cushing}, {Rayner}, \&
  {Vacca}}]{2005ApJ...623.1115C}
{Cushing}, M.~C., {Rayner}, J.~T., \& {Vacca}, W.~D. 2005, \apj, 623, 1115

\bibitem[{{Dupuy} {et~al.}(2020){Dupuy}, {Liu}, {Magnier}, {Best}, {Baraffe},
  {Chabrier}, {Forveille}, {Metchev}, \& {Tremblin}}]{2020RNAAS...4...54D}
{Dupuy}, T.~J., {Liu}, M.~C., {Magnier}, E.~A., {et~al.} 2020, Research Notes
  of the American Astronomical Society, 4, 54

\bibitem[{{Faherty} {et~al.}(2016){Faherty}, {Riedel}, {Cruz}, {Gagne},
  {Filippazzo}, {Lambrides}, {Fica}, {Weinberger}, {Thorstensen}, {Tinney},
  {Baldassare}, {Lemonier}, \& {Rice}}]{2016ApJS..225...10F}
{Faherty}, J.~K., {Riedel}, A.~R., {Cruz}, K.~L., {et~al.} 2016, \apjs, 225, 10

\bibitem[{{Filippazzo} {et~al.}(2015){Filippazzo}, {Rice}, {Faherty}, {Cruz},
  {Van Gordon}, \& {Looper}}]{2015ApJ...810..158F}
{Filippazzo}, J.~C., {Rice}, E.~L., {Faherty}, J., {et~al.} 2015, \apj, 810,
  158

\bibitem[{{Freudling} {et~al.}(2013){Freudling}, {Romaniello}, {Bramich},
  {Ballester}, {Forchi}, {Garc{\'\i}a-Dabl{\'o}}, {Moehler}, \&
  {Neeser}}]{2013A&A...559A..96F}
{Freudling}, W., {Romaniello}, M., {Bramich}, D.~M., {et~al.} 2013, \aap, 559,
  A96

\bibitem[{{Gauza} {et~al.}(2015){Gauza}, {B{\'e}jar}, {P{\'e}rez-Garrido},
  {Zapatero Osorio}, {Lodieu}, {Rebolo}, {Pall{\'e}}, \& {Nowak}}]{gauza2015}
{Gauza}, B., {B{\'e}jar}, V. J.~S., {P{\'e}rez-Garrido}, A., {et~al.} 2015,
  \apj, 804, 96

\bibitem[{{Hinkley} {et~al.}(2022){Hinkley}, {Carter}, {Ray}, {Skemer},
  {Biller}, {Choquet}, {Millar-Blanchaer}, {Sallum}, {Miles}, {Whiteford},
  {Patapis}, {Perrin}, {Pueyo}, {Schneider}, {Stapelfeldt}, {Wang},
  {Ward-Duong}, {Bowler}, {Boccaletti}, {Girard}, {Hines}, {Kalas}, {Kammerer},
  {Kervella}, {Leisenring}, {Pantin}, {Zhou}, {Meyer}, {Liu}, {Bonnefoy},
  {Currie}, {McElwain}, {Metchev}, {Wyatt}, {Absil}, {Adams}, {Barman},
  {Baraffe}, {Bonavita}, {Booth}, {Bryan}, {Chauvin}, {Chen}, {Danielski}, {De
  Furio}, {Factor}, {Fortney}, {Grady}, {Greenbaum}, {Henning}, {Janson},
  {Kennedy}, {Kenworthy}, {Kraus}, {Kuzuhara}, {Lagage}, {Lagrange},
  {Launhardt}, {Lazzoni}, {Lloyd}, {Marino}, {Marley}, {Martinez}, {Marois},
  {Matthews}, {Matthews}, {Mawet}, {Phillips}, {Petrus}, {Quanz},
  {Quirrenbach}, {Rameau}, {Rebollido}, {Rickman}, {Samland}, {Sargent},
  {Schlieder}, {Sivaramakrishnan}, {Stone}, {Tamura}, {Tremblin}, {Uyama},
  {Vasist}, {Vigan}, {Wagner}, \& {Ygouf}}]{2022arXiv220512972H}
{Hinkley}, S., {Carter}, A.~L., {Ray}, S., {et~al.} 2022, arXiv e-prints,
  arXiv:2205.12972

\bibitem[{{Kausch} {et~al.}(2015){Kausch}, {Noll}, {Smette}, {Kimeswenger},
  {Barden}, {Szyszka}, {Jones}, {Sana}, {Horst}, \&
  {Kerber}}]{2015A&A...576A..78K}
{Kausch}, W., {Noll}, S., {Smette}, A., {et~al.} 2015, \aap, 576, A78

\bibitem[{{Konopacky} {et~al.}(2013){Konopacky}, {Barman}, {Macintosh}, \&
  {Marois}}]{2013Sci...339.1398K}
{Konopacky}, Q.~M., {Barman}, T.~S., {Macintosh}, B.~A., \& {Marois}, C. 2013,
  Science, 339, 1398

\bibitem[{{Lucas} {et~al.}(2001){Lucas}, {Roche}, {Allard}, \&
  {Hauschildt}}]{2001MNRAS.326..695L}
{Lucas}, P.~W., {Roche}, P.~F., {Allard}, F., \& {Hauschildt}, P.~H. 2001,
  \mnras, 326, 695

\bibitem[{{Manjavacas} {et~al.}(2018){Manjavacas}, {Apai}, {Zhou}, {Karalidi},
  {Lew}, {Schneider}, {Cowan}, {Metchev}, {Miles-P{\'a}ez}, {Burgasser},
  {Radigan}, {Bedin}, {Lowrance}, \& {Marley}}]{2018AJ....155...11M}
{Manjavacas}, E., {Apai}, D., {Zhou}, Y., {et~al.} 2018, \aj, 155, 11

\bibitem[{{Manjavacas} {et~al.}(2020){Manjavacas}, {Lodieu}, {B{\'e}jar},
  {Zapatero-Osorio}, {Boudreault}, \& {Bonnefoy}}]{2020MNRAS.491.5925M}
{Manjavacas}, E., {Lodieu}, N., {B{\'e}jar}, V.~J.~S., {et~al.} 2020, \mnras,
  491, 5925

\bibitem[{{Martin} {et~al.}(2017){Martin}, {Mace}, {McLean}, {Logsdon}, {Rice},
  {Kirkpatrick}, {Burgasser}, {McGovern}, \& {Prato}}]{2017ApJ...838...73M}
{Martin}, E.~C., {Mace}, G.~N., {McLean}, I.~S., {et~al.} 2017, \apj, 838, 73

\bibitem[{{Metchev} {et~al.}(2015){Metchev}, {Heinze}, {Apai}, {Flateau},
  {Radigan}, {Burgasser}, {Marley}, {Artigau}, {Plavchan}, \&
  {Goldman}}]{2015ApJ...799..154M}
{Metchev}, S.~A., {Heinze}, A., {Apai}, D., {et~al.} 2015, \apj, 799, 154

\bibitem[{{Miles} {et~al.}(2018){Miles}, {Skemer}, {Barman}, {Allers}, \&
  {Stone}}]{2018ApJ...869...18M}
{Miles}, B.~E., {Skemer}, A.~J., {Barman}, T.~S., {Allers}, K.~N., \& {Stone},
  J.~M. 2018, \apj, 869, 18

\bibitem[{{Modigliani} {et~al.}(2010){Modigliani}, {Goldoni}, {Royer},
  {Haigron}, {Guglielmi}, {Fran{\c{c}}ois}, {Horrobin}, {Bristow}, {Vernet},
  {Moehler}, {Kerber}, {Ballester}, {Mason}, \&
  {Christensen}}]{2010SPIE.7737E..28M}
{Modigliani}, A., {Goldoni}, P., {Royer}, F., {et~al.} 2010, in Society of
  Photo-Optical Instrumentation Engineers (SPIE) Conference Series, Vol. 7737,
  Observatory Operations: Strategies, Processes, and Systems III, ed. D.~R.
  {Silva}, A.~B. {Peck}, \& B.~T. {Soifer}, 773728

\bibitem[{{Molli{\`e}re} {et~al.}(2022){Molli{\`e}re}, {Molyarova}, {Bitsch},
  {Henning}, {Schneider}, {Kreidberg}, {Eistrup}, {Burn}, {Nasedkin},
  {Semenov}, {Mordasini}, {Schlecker}, {Schwarz}, {Lacour}, {Nowak}, \&
  {Schulik}}]{2022arXiv220413714M}
{Molli{\`e}re}, P., {Molyarova}, T., {Bitsch}, B., {et~al.} 2022, arXiv
  e-prints, arXiv:2204.13714

\bibitem[{{Molli{\`e}re} {et~al.}(2020){Molli{\`e}re}, {Stolker}, {Lacour},
  {Otten}, {Shangguan}, {Charnay}, {Molyarova}, {Nowak}, {Henning}, {Marleau},
  {Semenov}, {van Dishoeck}, {Eisenhauer}, {Garcia}, {Garcia Lopez}, {Girard},
  {Greenbaum}, {Hinkley}, {Kervella}, {Kreidberg}, {Maire}, {Nasedkin},
  {Pueyo}, {Snellen}, {Vigan}, {Wang}, {de Zeeuw}, \&
  {Zurlo}}]{2020A&A...640A.131M}
{Molli{\`e}re}, P., {Stolker}, T., {Lacour}, S., {et~al.} 2020, \aap, 640, A131

\bibitem[{{Naud} {et~al.}(2014){Naud}, {Artigau}, {Malo}, {Albert}, {Doyon},
  {Lafreni{\`e}re}, {Gagn{\'e}}, {Saumon}, {Morley}, {Allard}, {Homeier},
  {Beichman}, {Gelino}, \& {Boucher}}]{2014ApJ...787....5N}
{Naud}, M.-E., {Artigau}, {\'E}., {Malo}, L., {et~al.} 2014, \apj, 787, 5

\bibitem[{{Petrus} {et~al.}(2020){Petrus}, {Bonnefoy}, {Chauvin}, {Babusiaux},
  {Delorme}, {Lagrange}, {Florent}, {Bayo}, {Janson}, {Biller}, {Manjavacas},
  {Marleau}, \& {Kopytova}}]{2020AA...633A.124P}
{Petrus}, S., {Bonnefoy}, M., {Chauvin}, G., {et~al.} 2020, \aap, 633, A124

\bibitem[{{Petrus} {et~al.}(2021){Petrus}, {Bonnefoy}, {Chauvin}, {Charnay},
  {Marleau}, {Gratton}, {Lagrange}, {Rameau}, {Mordasini}, {Nowak}, {Delorme},
  {Boccaletti}, {Carlotti}, {Houll{\'e}}, {Vigan}, {Allard}, {Desidera},
  {D'Orazi}, {Hoeijmakers}, {Wyttenbach}, \& {Lavie}}]{2021AA...648A..59P}
{Petrus}, S., {Bonnefoy}, M., {Chauvin}, G., {et~al.} 2021, \aap, 648, A59

\bibitem[{{Saumon} \& {Marley}(2008)}]{2008ApJ...689.1327S}
{Saumon}, D. \& {Marley}, M.~S. 2008, \apj, 689, 1327

\bibitem[{{Skemer} {et~al.}(2014){Skemer}, {Marley}, {Hinz}, {Morzinski},
  {Skrutskie}, {Leisenring}, {Close}, {Saumon}, {Bailey}, {Briguglio},
  {Defrere}, {Esposito}, {Follette}, {Hill}, {Males}, {Puglisi}, {Rodigas}, \&
  {Xompero}}]{2014ApJ...792...17S}
{Skemer}, A.~J., {Marley}, M.~S., {Hinz}, P.~M., {et~al.} 2014, \apj, 792, 17

\bibitem[{{Smette} {et~al.}(2015){Smette}, {Sana}, {Noll}, {Horst}, {Kausch},
  {Kimeswenger}, {Barden}, {Szyszka}, {Jones}, {Gallenne}, {Vinther},
  {Ballester}, \& {Taylor}}]{2015A&A...576A..77S}
{Smette}, A., {Sana}, H., {Noll}, S., {et~al.} 2015, \aap, 576, A77

\bibitem[{{Stone} {et~al.}(2016){Stone}, {Skemer}, {Kratter}, {Dupuy}, {Close},
  {Eisner}, {Fortney}, {Hinz}, {Males}, {Morley}, {Morzinski}, \&
  {Ward-Duong}}]{2016ApJ...818L..12S}
{Stone}, J.~M., {Skemer}, A.~J., {Kratter}, K.~M., {et~al.} 2016, \apjl, 818,
  L12

\bibitem[{{Su{\'a}rez} \& {Metchev}(2022)}]{2022MNRAS.513.5701S}
{Su{\'a}rez}, G. \& {Metchev}, S. 2022, \mnras, 513, 5701

\bibitem[{{Tremblin} {et~al.}(2016){Tremblin}, {Amundsen}, {Chabrier},
  {Baraffe}, {Drummond}, {Hinkley}, {Mourier}, \&
  {Venot}}]{2016ApJ...817L..19T}
{Tremblin}, P., {Amundsen}, D.~S., {Chabrier}, G., {et~al.} 2016, \apjl, 817,
  L19

\bibitem[{{Tremblin} {et~al.}(2015){Tremblin}, {Amundsen}, {Mourier},
  {Baraffe}, {Chabrier}, {Drummond}, {Homeier}, \&
  {Venot}}]{2015ApJ...804L..17T}
{Tremblin}, P., {Amundsen}, D.~S., {Mourier}, P., {et~al.} 2015, \apjl, 804,
  L17

\bibitem[{{Tremblin} {et~al.}(2019){Tremblin}, {Padioleau}, {Phillips},
  {Chabrier}, {Baraffe}, {Fromang}, {Audit}, {Bloch}, {Burgasser}, {Drummond},
  {Gonz{\'a}lez}, {Kestener}, {Kokh}, {Lagage}, \&
  {Stauffert}}]{2019ApJ...876..144T}
{Tremblin}, P., {Padioleau}, T., {Phillips}, M.~W., {et~al.} 2019, \apj, 876,
  144

\bibitem[{{Tremblin} {et~al.}(2020){Tremblin}, {Phillips}, {Emery}, {Baraffe},
  {Lew}, {Apai}, {Biller}, \& {Bonnefoy}}]{2020A&A...643A..23T}
{Tremblin}, P., {Phillips}, M.~W., {Emery}, A., {et~al.} 2020, \aap, 643, A23

\bibitem[{{Venot} {et~al.}(2012){Venot}, {H{\'e}brard}, {Ag{\'u}ndez},
  {Dobrijevic}, {Selsis}, {Hersant}, {Iro}, \&
  {Bounaceur}}]{2012A&A...546A..43V}
{Venot}, O., {H{\'e}brard}, E., {Ag{\'u}ndez}, M., {et~al.} 2012, \aap, 546,
  A43

\bibitem[{{Vernet} {et~al.}(2011){Vernet}, {Dekker}, {D'Odorico}, {Kaper},
  {Kjaergaard}, {Hammer}, {Randich}, {Zerbi}, {Groot}, {Hjorth}, {Guinouard},
  {Navarro}, {Adolfse}, {Albers}, {Amans}, {Andersen}, {Andersen}, {Binetruy},
  {Bristow}, {Castillo}, {Chemla}, {Christensen}, {Conconi}, {Conzelmann},
  {Dam}, {de Caprio}, {de Ugarte Postigo}, {Delabre}, {di Marcantonio},
  {Downing}, {Elswijk}, {Finger}, {Fischer}, {Flores}, {Fran{\c{c}}ois},
  {Goldoni}, {Guglielmi}, {Haigron}, {Hanenburg}, {Hendriks}, {Horrobin},
  {Horville}, {Jessen}, {Kerber}, {Kern}, {Kiekebusch}, {Kleszcz}, {Klougart},
  {Kragt}, {Larsen}, {Lizon}, {Lucuix}, {Mainieri}, {Manuputy}, {Martayan},
  {Mason}, {Mazzoleni}, {Michaelsen}, {Modigliani}, {Moehler}, {M{\o}ller},
  {Norup S{\o}rensen}, {N{\o}rregaard}, {P{\'e}roux}, {Patat}, {Pena}, {Pragt},
  {Reinero}, {Rigal}, {Riva}, {Roelfsema}, {Royer}, {Sacco}, {Santin},
  {Schoenmaker}, {Spano}, {Sweers}, {Ter Horst}, {Tintori}, {Tromp}, {van
  Dael}, {van der Vliet}, {Venema}, {Vidali}, {Vinther}, {Vola}, {Winters},
  {Wistisen}, {Wulterkens}, \& {Zacchei}}]{2011A&A...536A.105V}
{Vernet}, J., {Dekker}, H., {D'Odorico}, S., {et~al.} 2011, \aap, 536, A105

\bibitem[{{Zhou} {et~al.}(2020){Zhou}, {Bowler}, {Morley}, {Apai}, {Kataria},
  {Bryan}, \& {Benneke}}]{2020AJ....160...77Z}
{Zhou}, Y., {Bowler}, B.~P., {Morley}, C.~V., {et~al.} 2020, \aj, 160, 77

\end{thebibliography}

\begin{appendix}


\section{X-shooter Observing Log}
\label{Appendix:obs_log}

\begin{table*}[ht!]
\caption{Observing log}
\label{tab:obs}
\renewcommand{\arraystretch}{1.3}
\begin{center}
\small
\begin{tabular}{l|lllllll}
\hline
\hline
Seq 	&	Date	    &	UT Start-Time   &  $\mathrm{DIT}$   & $\mathrm{NDIT}$   & $\mathrm{NEXP}$   & $<$Seeing$>$  &  Airmass	 	\\
		& (yyyy-mm-dd)	&			(hh:mm)	&	(s)				&				    &				    &      (")      &	            \\
\hline
1       & 2018-05-28    & 01:36             & 670/670/234       &  1/1/3            & 12/12/12          & 0.68          & 1.05      \\
2       & 2018-05-28    & 02:36             & 670/670/234       &  1/1/3            & 12/12/12          & 0.74          & 1.17      \\

\hline
\hline
\end{tabular}
\end{center}
\tablefoot{The seeing is measured at 0.5~$\muup$m and given for the visible arm. The DIT {(Detector Integration Time)} values refer to the individual exposure time per frame in the UVB, VIS, and NIR arms respectively. NDIT are the number of individual frames per exposure, and $N_{EXP}$  the number of exposures  in the UVB, VIS, and NIR arms.}\\
\end{table*}

The Table \ref{tab:obs} reports the observing conditions during our data acquirement.

\section{Empirical analysis}
\label{App:emp_ana}
\begin{figure}[ht!]
\centering
\includegraphics[width=1\hsize]{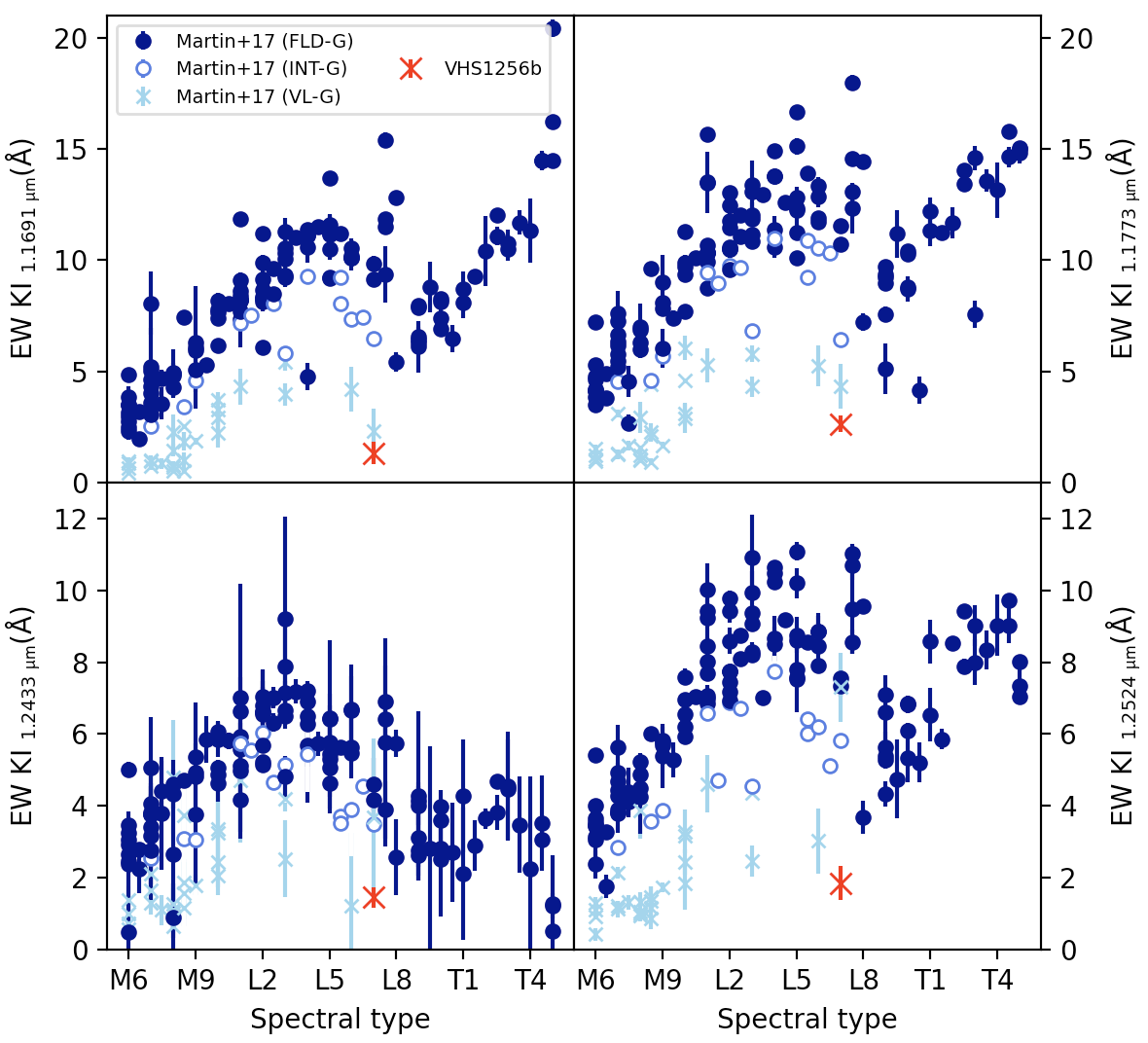}
    \caption{Equivalent widths of the detected K\,I lines of VHS\,1256\,b compared to the values of the sample of old (FLD-G) and young, low gravity (VL-G) and intermediate gravity (INT-G), brown dwarfs of \cite{2017ApJ...838...73M}.}
    \label{fig:empirical}
\end{figure}

The medium spectral resolution of our data allows us to identify several strong absorption lines between 0.8 and 1.3~\mic, like the neutral cesium (Cs, 0.852~\mic), and the neutral potassium (K I) doublets (1.168, 1.177~\mic~ and 1.243, 1.254~\mic). The sodium (Na I) doublet (1.138 and 1.141~\mic) is also detected. These lines were not resolved in the early study of \cite{gauza2015} given the lower spectral resolution of the GTC/OSIRIS optical and NTT/SofI nIR spectra. We therefore derived the equivalent width (EW) of the KI lines and compared them with the ones obtained by \cite{2017ApJ...838...73M} for a sample of old and young (low gravity and intermediate gravity) brown dwarfs as shown in Fig.\,\ref{fig:empirical}. 

The EW of alkaline lines in the NIR have been shown to correlate with surface gravity \citep{2005ApJ...623.1115C, 2013ApJ...772...79A, 2014AA...562A.127B, 2020MNRAS.491.5925M}. The strength of these lines increases with the surface gravity due to a higher pressure in the photosphere. Therefore, they are great tracers of age. To calculate these EWs, we have exploited the resolution of X-Shooter by applying the following procedure. We constructed 100 spectra whose flux at each wavelength is randomly generated with a Gaussian law of mean the initial data value and standard deviation the data error. For each spectrum we estimated the pseudo-continuum at the position of the line by subtracting a Lorentzian law, fitted on the line, to the spectrum, which is then degrade to a resolution R$_{\lambda}$\,=\,250. The EW is calculated with:

    \begin{equation}
    \label{Eq:EW}
    EW~=~\sum_{i=1}^{n}~ \biggr[ 1 - \frac{f(\lambda_{i})}{f_{c}(\lambda_{i})} \biggl]\Delta \lambda_{i}
    \end{equation}  
 
with $f(\lambda_{i})$ and $f_{c}(\lambda_{i})$ the flux of the line and the pseudo-continuum, respectively, for each wavelength $\lambda_{i}$. $\Delta \lambda_{i}$ is the wavelength step. We estimated the final EW from the mean of the serie of these 100 EWs and the final error from its standard deviation. 

The Figure \ref{fig:empirical} compares the EW of each K I lines calculated for VHS1256 b and the EWs calculated by \cite{2017ApJ...838...73M} who used the method from \cite{2013ApJ...772...79A}, more appropriate for their data (R$_{\lambda}\sim$~2000). For each K\,I line, the EW calculated for VHS1256 b is substantially weaker that the one of field and intermediate-gravity (INT-G) which confirms the VL-G classification and young age determined by \cite{gauza2015} based on the H$_2$OD and $H$-continuum spectral indices. 

\section{Grid description}
\label{Appendix:grids}
In this Appendix we will briefly describe the main properties of the \texttt{ATMO} \citep{2015ApJ...804L..17T} models and the grids that they generated and that have been used in our study (see the Table \ref{tab:grid_param}). These grids are publicly available at \url{https://opendata.erc-atmo.eu}. The \texttt{ATMO} models assume that clouds are not needed to reproduce the shape of the SED of brown dwarfs (apart from the 10-\mic silicate feature). \cite{2016ApJ...817L..19T} have proposed that diabatic convective processes \citep{2019ApJ...876..144T} induced by out-of-equilibrium chemistry of CO/CH$_{4}$ and N$_{2}$/NH$_{3}$ can reduce the temperature gradient in the atmosphere and reproduce the reddening previously thought to occur by clouds. The grids assume 
a modification of the temperature gradient with an effective adiabatic index. 
The levels modified are in between 2 and 200 bars at log(g)\,=\,5.0 and are scaled by $\times 10^{\mathrm{log(g)}-5}$ at other surface gravities. Out-of-equilirium chemistry is used with $k_{zz} = 10^5$ cm$^2$/s at log(g)\,=\,5.0 and are scaled by $\times 10^{2(5-\mathrm{log(g)})}$ at other surface gravities. The mixing length is assumed to be 2 scale height at 20 bars and higher pressures at log(g)\,=\,5.0 and is scaled down by the ratio between the local pressure and the pressure at 20 bars for lower pressures. The 20 bars limit is scaled by $\times 10^{\mathrm{log(g)}-5}$ at other surface gravities. The chemistry includes 277 species and out-of-equilibrium chemistry has been performed using the model of \citet{2012A&A...546A..43V}, rainout is assumed to occur for species that are not included in the out-of-equilibrium model. Opacity sources include H$_2$-H$_2$, H$_2$-He, H$_2$O, CO$_2$, CO, CH$_4$, NH$_3$, Na, K, Li, Rb, Cs, TiO, VO, FeH, PH$_3$, H$_2$S, HCN, C$_2$H$_2$, SO$_2$, Fe, H$^-$, and the Rayleigh scattering opacities for H$_2$, He, CO, N$_2$, CH$_4$, NH$_3$, H$_2$O, CO$_2$, H$_2$S, SO$_2$. The grids explore the following parameters: effective temperatures between 500 and 3000K (step 100K); log(g) between 2.5 and 5.5 (step 0.5); effective adiabatic index (reddenning) with three values 1.05, 1.03, 1.01; metallicity with five values -0.6, -0.3, 0, +0.3, +0.6; C/O ratio with three values 0.3, solar, 0.7. The last version has been used to generate a grid of synthetic spectra at medium resolution (R$_{\lambda}\sim$10000) from 0.83 to 2.5~\mic as well as lower resolution synthetic spectra (R$_{\lambda}\sim$1000) between 0.8 and 20~\mic. 

\begin{table}[h]
\caption{Parameters space of the grid of synthetic spectra \texttt{ATMO}. \Teff~is the effective temperature, log(g) is the surface gravity, [M/H] is the metallicity, C/O is the carbon-oxygen ratio, and $\mathrm{\gamma}$ is the reduced adiabatic index.}
\label{tab:grid_param}
\renewcommand{\arraystretch}{1.31}
\begin{center}
\small
\begin{tabular}{l|ll}
\hline
\hline
     	                            &	        &     \texttt{ATMO}    \\
\hline
\multirow{2}{*}{\Teff~(K)}          & Range     &	  [800, 3000]    \\
                                    & Step      &     50  \\
\hline
\multirow{2}{*}{log(g) (dex)}       & Range     &	  [2.5, 5.5]    \\
                                    & Step      &	  0.5    \\
\hline
\multirow{2}{*}{[M/H]}              & Range     &	  [-0.6, 0.6]    \\
                                    & Step      &	  0.3    \\
\hline
\multirow{2}{*}{C/O}             & Range     &	  [0.3, 0.7]    \\
                                    & Step      &   0.25    \\
\hline
\multirow{2}{*}{$\mathrm{\gamma}$}  & Range     &   [1.01, 1.05]    \\
                                    & Step      &   0.02    \\
\hline
\hline
\end{tabular}
\end{center}
\end{table}

\section{Estimates of atmospheric and dynamic parameters}

\begin{table*}[h]
\caption{Atmospheric parameters of VHS 1256-1257 b inferred from different wavelength ranges, using \texttt{ForMoSA} and the \texttt{ATMO} model grid. NC is for "No Constraint". The error bars given here are purely statistical and derived from the propagation of the small error bars of our data through the Bayesian inversion. The method used to estimate the final values is described in the Appendix \ref{App:Adopt_val}.}
\label{tab:fits_param}
\renewcommand{\arraystretch}{2.1}
\begin{center}
\small
\begin{tabular}{c|ccccccccc}
\hline
\hline
$\mathrm{\lambda_{min}-\lambda_{max}}$               & \Teff	                & log(g)                   & [M/H]                     & C/O                       & $\mathrm{\gamma}$         & R                         & RV                        & v\,sin(i)     & log(L/L$_{\odot}$) \\
 (\mic)                      & (K)	                & (dex)                     &                           &                           &                           & (\Rjup)                   & (km\,s$^{-1}$)             & (km\,s$^{-1}$) &    (dex) \\
\hline
1.100--2.480 (JHK)   & 1326$^{+2}_{-2}$      & 3.75$^{+0.02}_{-0.02}$	& 0.45$^{+0.03}_{-0.02}$    & 0.63$^{+0.01}_{-0.01}$	& 1.013$^{+0.001}_{-0.001}$	& 0.91$^{+0.01}_{-0.01}$    & 4.47$^{+0.39}_{-0.38}$    & ---               & $-4.63^{+0.01}_{-0.01}$  \\
0.670--1.100 (RI)    & 1361$^{+4}_{-3}$	& 5.49$^{+0.01}_{-0.01}$	& 0.42$^{+0.01}_{-0.02}$	& > 0.69	                & < 1.02	                    & 0.50$^{+0.01}_{-0.01}$	                    & 6.04$^{+1.36}_{-0.59}$	& ---	            	& $-5.10^{+0.01}_{-0.01}$	\\
1.100--1.356 (J)     & 1414$^{+6}_{-21}$	& 3.90$^{+0.09}_{-0.08}$	& 0.44$^{+0.07}_{-0.07}$	& > 0.68	                & 1.02$^{+0.01}_{-0.01}$	                    & 0.72$^{+0.03}_{-0.01}$	                    & 3.32$^{+0.75}_{-0.82}$	& ---	           	& $-4.73^{+0.01}_{-0.01}$	\\
1.410--1.810 (H)     & 1390$^{+6}_{-7}$	& 3.50$^{+0.02}_{-0.01}$	& -0.05$^{+0.04}_{-0.03}$	& > 0.66	                & 1.04$^{+0.01}_{-0.01}$	                    & 0.82$^{+0.01}_{-0.01}$	                    & 3.80$^{+0.43}_{-0.61}$	& ---	           	& $-4.64^{+0.01}_{-0.01}$	\\
1.952--2.478 (K)     & 1417$^{+17}_{-17}$	& 4.02$^{+0.05}_{-0.05}$	& 0.08$^{+0.05}_{-0.05}$	& > 0.68	                & 1.02$^{+0.01}_{-0.01}$	                    & 0.81$^{+0.01}_{-0.01}$	                    & 1.60$^{+1.13}_{-0.92}$	& ---	           	& $-4.62^{+0.01}_{-0.01}$	\\
2.90--4.14 (L$_{P}$)$^a$     & 1145$^{+13}_{-10}$	& < 2.69	& 0.47$^{+0.08}_{-0.09}$	& 0.39$^{+0.06}_{-0.05}$	                & 1.03$^{+0.01}_{-0.01}$	                    & 1.25$^{+0.02}_{-0.02}$	                    & 5.62$^{+3.15}_{-3.25}$	& ---	           	& $-4.61^{+0.01}_{-0.01}$	\\
\hline
1.12--1.65 (HST min)    & 1309$^{+3}_{-3}$	& 3.34$^{+0.10}_{-0.11}$	& NC	& 	   NC             &	    < 1.02               &  --- & --- & --- & ---\\    
1.12--1.65 (HST max)    & 1336$^{+5}_{-3}$	& 3.40$^{+0.08}_{-0.09}$	& NC	& 	   NC             &	    < 1.02               &  --- & --- & --- & --- \\    
1.12--1.65 (X-Shooter)    & 1331$^{+2}_{-2}$	& 3.70$^{+0.03}_{-0.03}$	& NC	& 	   NC             &	    < 1.02               &  --- & --- & --- & --- \\    
\hline
2.280--2.410 ($^{12}$CO)    & 1479$^{+72}_{-63}$	& 4.21$^{+0.18}_{-0.18}$	& 0.02$^{+0.11}_{-0.10}$	& > 0.63	                & < 1.02	                    & ---	                    & 4.04$^{+1.27}_{-1.23}$	& ---	           	& ---	\\
1.225--1.275 (K I)$^b$   & 1573$^{+36}_{-46}$	& 4.25$^{+0.20}_{-0.21}$	& > 0.32	                & 0.43$^{+0.11}_{-0.09}$	& < 1.015	                    & ---	                    & $-1.32^{+4.33}_{-3.26}$   & < 37	         & ---	\\
\hline
\hline
Adopted values   & 1380$\pm$54	& 3.97$\pm$0.48	& 0.21$\pm$0.29 & > 0.63	& < 1.02	                    & 0.99$\pm$0.28	                    &  1.41$\pm$5.99  & < 37	         & 4.67$\pm$0.07	\\
\hline
\hline
\end{tabular}
\end{center}
\footnotesize{$^a$ Low resolution version of the \texttt{ATMO} model}\\
\footnotesize{$^b$ With continuum subtracted}\\
\end{table*}

\label{App:Adopt_val}
As illustrated in the Table \ref{tab:fits_param}, the estimate of each parameter is strongly dependent on the wavelength range used for the fit. To take into account these variations and propose a conservative estimate of the atmospheric parameters of VHS1256\,b, we followed the method described in \cite{2020AA...633A.124P}:

\begin{itemize}
    \item The final \Teff~is mainly constrained by the low-resolution spectral information. It is defined by the extreme estimates from the fits that used the $JHK$-, $J$-, $H$-, and $K$-bands.    
    \item The final log(g)~is constrained by the low-resolution spectral information, but also by the molecular and atomic absorption. It is defined by the extreme estimates from the fits that used the [1.225–1.275]~\mic wavelength range, and the $JHK$-, $J$-, $H$-, and $K$-bands.
    \item Two modes seem to be observed for the metallicity. One solar ($\sim$0.0) and one super-solar ($\sim$0.5). That is why we have chosen to let it unconstrained by defining its final value based on the extreme estimates from all of the fits.
    \item For most of the fits, the C/O converges to the upper edge of the grid. We defined its adopted values as the most conservative lower limit.
    \item For most of the fits, the $\gamma$ converges to the lower edge of the grid. We defined its adopted values as the most conservative upper limit.
    \item Because the fit between 0.670 and 1.100~\mic~does not seem to reproduce the data, it has been excluded to estimate the radius. Its adopted value is based on the extreme estimates from all of the other fits.
    \item The adopted value of the radial velocity is defined as the extreme estimates from all of the fits.
    \item For the same reason as for the radius, we excluded the fit on the $RI$-band to estimate the bolometric luminosity. Its adopted value is defined as the extreme estimates from all of the other fits.
\end{itemize}

\end{appendix}

\end{document}